# Quantum Information Science: Emerging No More

*Carlton M. Caves*

Center for Quantum Information and Control, University of New Mexico, Albuquerque, New Mexico 87131-0001, USA

Final: 2013 January 30

Quantum information science (QIS) is a new field of enquiry, nascent in the 1980s, founded firmly in the 1990s, exploding in the 2010s, now established as a discipline for the 21st Century.

Born in obscurity, then known as the foundations of quantum mechanics, the field began in the 60s and 70s with studies of Bell inequalities. These showed that the predictions of quantum mechanics cannot be squared with the belief, called local realism, that physical systems have realistic properties whose pre-existing values are revealed by measurements. The predictions of quantum mechanics for separate systems, correlated in the quantum way that we now call entanglement, are at odds with any version of local realism. Experiments in the early 80s demonstrated convincingly that the world comes down on the side of quantum mechanics. With local realism tossed out the window, it was natural to dream that quantum correlations could be used for faster-than-light communication, but this speculation was quickly shot down, and the shooting established the principle that quantum states cannot be copied.

A group consisting of quantum opticians, electrical engineers, and mathematical physicists spent the 60s and 70s studying quantum measurements, getting serious about what can be measured and how well, going well beyond the description of observables that was (and often still is) taught in quantum-mechanics courses. This was not an empty exercise: communications engineers needed a more general description of quantum measurements to describe communications channels and to assess their performance. These developments led, by the early 80s, to a general formulation of quantum dynamics, capable of describing all the state changes permitted by quantum mechanics, including the dynamics of open quantum systems and the state transformations associated with the most general measurements. An important advance was a quantitative understanding of the inability to determine reliably the quantum state of a single system from measurements.

The 80s spawned several key ideas. A major discovery was quantum key distribution, the ability to distribute secret keys to distant parties. The keys can be used to encode messages for secure communication between the parties, conventionally called Alice and Bob, with the security guaranteed by quantum mechanics. In addition, early in the decade, physicists and computer scientists began musing that the dynamics of quantum systems might be a form of information processing. Powerful processing it would be, since quantum dynamics is difficult to simulate, difficult because when many quantum systems interact, the number of probability amplitudes grows exponentially with the number of systems. Unlike probabilities, one cannot simulate the evolution of the amplitudes by tracking underlying local realistic properties that undergo probabilistic transitions: the interference of probability amplitudes forbids; there are no underlying properties. If quantum systems are naturally doing information processing that can't be easily simulated, then perhaps they can be turned to doing information-processing jobs for us. So David Deutsch suggested in the mid-80s, and thus was born the quantum computer.

As the 90s dawned, two new capabilities emerged. The first, entanglement-based quantum key distribution, relies for security on the failure of local realism, which says that there is no shared key till

Alice and Bob observe it. This turns quantum entanglement and the associated failure of local realism from curiosities into a tool. The second capability, teleportation, lets the ubiquitous Alice and Bob, who share prior entanglement, transfer an arbitrary quantum state of a system at Alice's end to a system at Bob's end, at the cost of Alice's communicating a small amount of classical information to Bob. Surprising this is, because the state must be transferred without identifying it or copying it, both of which are forbidden. Sure enough, the classical bits that Alice sends to Bob bear no evidence of the state's identity, nor is any remnant of the state left at Alice's end. The correlations of pre-shared entanglement provide the magic that makes teleportation work.

These two protocols fed a growing belief that quantum mechanics is a framework describing information processing in quantum systems. The basic unit of this quantum information, called a qubit, is any two-level system. The general formulation of quantum dynamics provides the rules for preparing quantum systems, controlling and manipulating their evolution to perform information-processing tasks, and reading out the results as classical information.

The mid-90s brought a revolution, sparked by discoveries of what can be done in principle combining with laboratory advances in atomic physics and quantum optics that expanded what can be done in practice. The first discovery, from Peter Shor, was an efficient quantum algorithm for factoring integers, a task for which there is believed to be no efficient classical algorithm. The second was a proposal from Ignacio Cirac and Peter Zoller for a realistic quantum computer using trapped ions. This proposal drew on a steady stream of advances that promised the ability to control and manipulate individual neutral atoms or ions, all the while maintaining quantum coherence, and applied these to the design of the one- and two-qubit gates necessary for quantum computation. The third discovery, quantum error correction, was perhaps the most surprising and important finding about the nature of quantum mechanics since its formulation in the 1920s. Discovered independently by Peter Shor and Andrew Steane, quantum error correction (Fig. 1), when extended to techniques of fault-tolerant quantum computation, allows a quantum computer to compute indefinitely without error, provided the occurrence of errors is reduced below a threshold rate.

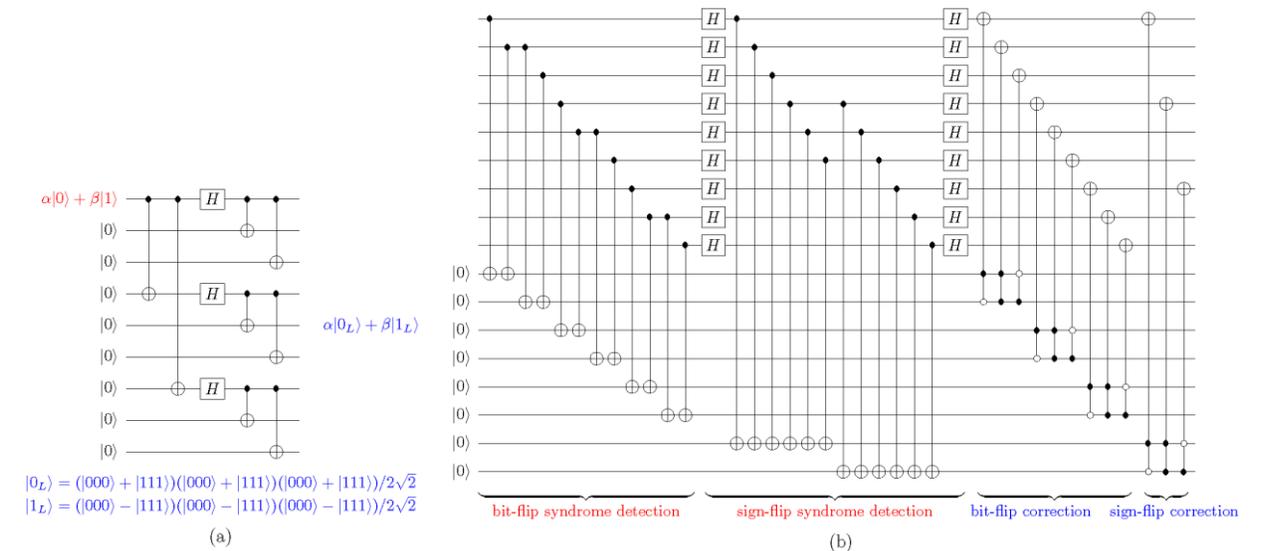

Figure 1. (a) Coding circuit for Shor nine-qubit quantum code. An arbitrary superposition of the 0 and 1 (physical) states of the top qubit is encoded into an identical superposition of the 0 and 1 (logical) states of nine qubits. (b) Error (syndrome) detection and error correction circuit for Shor nine-qubit code. Six ancilla qubits are used to detect a bit flip (exchange of 0 and 1) in any of the nine encoded qubits, and two ancilla qubits are used to detect a relative sign change between 0 and 1 in any of the nine encoded qubits. Correction operations repair the errors. The code detects and corrects all single-qubit errors on the encoded qubits and some multi-qubit errors.

Definitely a field by 2000, QIS galloped into the new millennium, an amalgam of researchers investigating the foundations of quantum mechanics, quantum opticians and atomic physicists building on a legacy of quantum coherence in atomic and optical systems, condensed-matter physicists working on implementing quantum logic in condensed systems, and a leavening of computer scientists bringing an information-theoretic perspective to all of quantum physics.

QIS researchers are implementing the fundamental processing elements for constructing a quantum computer in a variety of systems: ions trapped in electromagnetic fields, controlled by laser pulses and herded to interaction sites by electric fields (Fig. 2); circuit-QED, in which superconducting qubits are controlled by microwaves in cavities and transmission lines; neutral atoms cooled and trapped, interacting via cold collisions or by excitation to Rydberg levels; impurity atoms, vacancies, and quantum dots in semiconductor or other substrates, controlled electronically or photonically; and photonic qubits processed through complicated linear-optical interferometers, capable of implementing efficient quantum computation provided they are powered by single-photon sources and the photons can be counted efficiently. As experimenters develop these basic elements for quantum information processing, theorists integrate them into architectures for full-scale quantum computers, including quantum error correction to suppress the deleterious effects of noise and of unwanted couplings to the external world that destroy quantum coherence. An active research effort explores the space of quantum error-correcting codes to find optimal codes for fault-tolerant quantum computation.

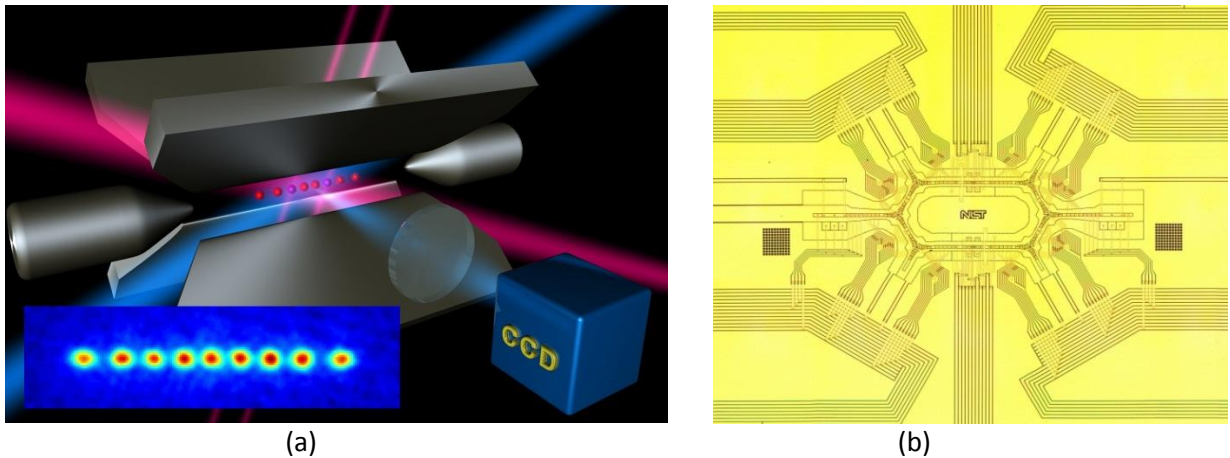

(a) (b)

Figure 2. (a) Linear ion trap. Ions (red) are trapped by a combination of DC and RF voltages. Two internal states of each ion, labeled 0 and 1, act as qubits. Laser beams (blue) drive quantum gate operations; two-qubit gates are mediated by the Coulomb repulsion between ions. Readout is by resonance fluorescence recorded by a CCD camera: absence or presence of fluorescence signals a qubit's 0 or 1 state. The inset shows detection of nine ions. Figure courtesy of R. Blatt, Quantum Optics and Spectroscopy Group, University of Innsbruck. (b) NIST Racetrack surface ion trap. Made of a quartz wafer coated with gold in an oval shape roughly two by four millimeters, this trap features 150 work zones, which are located just above the surface of the center ring structure and the six channels radiating out from its edge. Qubits are shuttled between zones, where they can be stored or manipulated for quantum information processing. The trap could be scaled up to a much larger number of zones. Figure courtesy of J. Amini, Ion Storage Group, National Institute of Standards and Technology.

Other researchers investigate exotic architectures for quantum computation, such as topological quantum computation, which encodes quantum information in many-body systems in a way that is naturally resistant to error, obviating or reducing the need for active quantum error correction. A prime candidate uses as qubits the quasiparticle excitations known as non-Abelian anyons, neither bosons nor fermions, but occurring naturally in fractional quantum-Hall states. Braiding of the anyons is used to realize quantum gates.

Experimenters verify the performance of quantum-information-processing devices using quantum-state and quantum-process tomography, techniques invented by quantum opticians to identify a quantum state when one can generate the same state over and over again.  The inefficiency of these tomographic techniques drives a search for more efficient ways to benchmark the performance of such devices.

Computer scientists explore the space of quantum algorithms, searching for algorithms that perform useful tasks more efficiently than can be done on a classical computer and seeking to understand generally the class of problems for which quantum computers provide an efficiency advantage.  One class of problems, present from the beginning of thinking about quantum computers, is the simulation of complex quantum systems, including complex materials, molecular structure, and the field theories of high-energy physics.

Quantum communications, the home of much early QIS thinking, now hosts the field's premier practical application, quantum key distribution.  Secret keys, distributed to distant parties over optical fiber and through free space, are used to encode messages for secure communication (Fig. 3).  Fundamental research continues on ensuring security in practical situations; using properties of the data exchanged in key distribution to guarantee security, instead of relying on an assumption that quantum mechanics is correct; the design of quantum repeaters, which, by using pre-shared entanglement, can extend the reach of key distribution beyond the usual limit set by losses in optical fiber; and the communication complexity of distributed information-processing tasks.

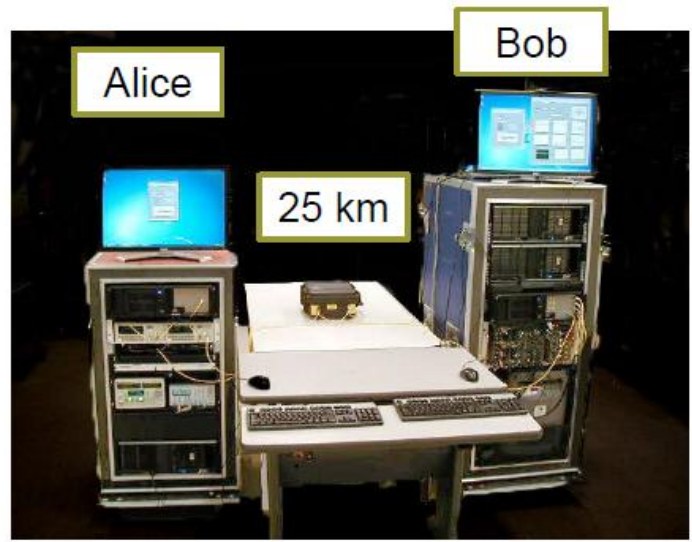

Figure 3.  Los Alamos prototype of a combined quantum-classical communications system that is used to secure electric-grid control data.  Alice, a quantum-communications transmitter and a classical optical transceiver, is connected to Bob, a quantum-communications receiver and a classical optical transceiver, through 25 km of optical fiber that carries both quantum and bi-directional classical communications.  Alice takes data needed to control the electric grid, encrypts it using shared key that is created by quantum key distribution over the same optical fiber, and transmits the data to Bob, where, decrypted, it is available for secure control of the grid.  Figure courtesy of the Los Alamos National Laboratory Quantum Communications Team, headed by J. E. Nordholt and R. J. Hughes.

The theory of entanglement is used in condensed-matter physics to characterize the ground and thermal states of many-body quantum systems with local interactions.  The degree and locality of entanglement become important variables for such systems, useful, for example, in characterizing when the low-energy states of the system can be efficiently described and simulated.

From its beginning, QIS has been a productive mixture of quantum weirdness and applications. The field has advanced by interplay between experiment and theory: experimental breakthroughs inspire theorists to dream of what might be, and the dreams of theorists inspire experimentalists to reduce the dreams to quantum reality. Physicists were forced to quantum mechanics, the highly successful framework for all of physical law, because the causal, deterministic, realistic narrative of classical physics fails for microscopic systems. Within the quantum framework, it is not surprising that one can do things that cannot be encompassed within a classical narrative; QIS is the discipline that does those things. In a broad sense, QIS is a sort of quantum engineering: though still rooted in fundamental science, QIS seeks ways to control the behavior of quantum systems and turn them to performing tasks we want done, instead of their doing what comes naturally.

QIS has burst well outside the bounds of what can be summarized in a potted history. To provide an illustration of what this means, I searched the web site of *Reviews of Modern Physics*, the premier journal for physics review articles, for all articles that have the phrase "quantum information" in the title or abstract. The search turned up 26 articles, the first of which appeared in 1999. These 26 articles collectively have 7,370 citations, 283 per article, and an h-index of 23. Promote the field to a full discipline.

There is more. Searching titles and abstracts misses many *RMP* articles associated with quantum information, so I searched the Tables of Contents of all issues of *RMP* from the 2000 to the end of 2012, adding to the previous list all those articles on quantum information that somehow neglected to include quantum information in the title or abstract, articles on the foundations of quantum mechanics, and articles on open quantum systems. This gives 44 review articles since 2000. In the period from 2000 to 2006, there were 16 articles, a rate of 2.6 per year. Since 2007, the pace has accelerated: there have been 28 review articles in *RMP*, a rate of 4.7 per year, more than one article per quarterly issue. And mind you, these are review articles, each of which cites dozens to hundreds of primary research papers.

It's time to stop talking about quantum information science as an ``emerging field.'' A discipline represented in every issue of *RMP* is no longer emerging. It has arrived.

**44 *RMP* articles from 2000 through the end of 2012 discussed in the text, plus the one mentioned article from 1999 and an additional classic from 1996. The 26 articles whose title or abstract contains the phrase "quantum information" are marked by underlining the first author's name.**